\documentclass[twocolumn,showpacs,preprintnumbers,prb]{revtex4}
\usepackage{amsmath}
\usepackage{dcolumn}
\usepackage{amssymb}
\usepackage{bm}
\usepackage{graphicx}
\usepackage{float}
\usepackage{subfig}
\usepackage{epstopdf}
\usepackage{float}
\newcommand{\abs}[1]{| #1 |} 
\newcommand{\ket}[1]{| #1 \rangle} 
\newcommand{\bra}[1]{\langle #1 |} 

\usepackage{color}

\begin{document}

\title{Quantum information transmission through a qubit chain with quasi-local dissipation}
\author{Roya Radgohar}
\affiliation{Department of Physics, Shiraz University, Shiraz, Iran}
\author{Laleh Memarzadeh}
\affiliation{Department of Physics, Sharif University of Technology, Teheran, Iran}
\author{Stefano Mancini}
\affiliation{School of Science and Technology, University of Camerino, I-62032 Camerino, Italy\\
INFN-Sezione di Perugia, I-06123 Perugia, Italy}
\date{\today}

\begin{abstract}

We study quantum information transmission in a Heisenberg-XY chain where qubits are affected by quasi-local environment action and compare it with the case of local action of the environment.
We find that for open boundary conditions the former situation always improves quantum state transfer process, especially for short chains. In contrast, for closed boundary conditions quasi-local environment results advantageous in the strong noise regime.
When the noise strength is comparable with the XY interaction strength, the state transfer fidelity through chain of odd/even number of qubits in presence of quasi-local environment results 
smaller/greater than that in presence of local environment.
\end{abstract}

\maketitle

\section{Introduction}\label{Introduction}
Quantum state transfer from one site to another is a key task in the field of quantum information and quantum computation  [\onlinecite{ref:CGLM}]. In addition to its essential role in quantum communication protocols, it is required for connecting small quantum processors in a quantum computer. Moreover, it might help us to get deep understanding of the behavior of natural systems, particularity biological systems [\onlinecite{ref:AJV}].
Transmitting information demands a physical system to serve as quantum channel, through which a quantum state is carried.
There exist some schemes considering qubits as the electronic states of trapped ions and transfer quantum information between ions through their Coulomb mutual interactions [\onlinecite{ref:CZ}], vibrational mode (bus-mode)         
 [\onlinecite{ref:SHRG}] or photons [\onlinecite{ref:BDMT}].
\par
Recently, Bose exploited nearest-neighbor interactions of Heisenberg-XY chain to perform swap operations to transfer quantum state from one end to another along the chain with some fidelity [\onlinecite{ref:BS}]. Despite the previous protocols [\onlinecite{ref:M}] applying external controls (which are decoherence sources as well) to switch coupling between qubits to transmit information, this protocol does not need any external control on the interconnecting qubits between the input and output qubits. An experimental implementation of the protocol based on Josephson junction array is provided in [\onlinecite{ref:RFB}]. Then, Datta in [\onlinecite{ref:CDEL}] proved that fixed but different couplings can provide perfect state transfer in Heisenberg-XY chain. 
\par Considering the unavoidable noisy effects in the dynamics of real quantum systems as well as the necessities of miniaturizing devices applied in quantum technologies like quantum computers, it may happen that nearest-neighbor qubits in Heisenberg-XY chain become so closely spaced to experience the same environment effects [\onlinecite{ref:Guo}]. Inspired by this fact, we would like to investigate quantum state transfer in a Heisenberg-XY chain with open and periodic boundary conditions in the presence of ``chained" (quasi-local) environments and compare the results with the case of local environments. This investigation may address the question of whether one should allow the qubits interact through chained environments or realize a situation in which each qubit interacts with its own environment. 
\par
Here, we found that chained environments in comparison with local ones, giving rise to indirect interactions between contaminated qubits, facilitates information transfer over a (short) chain with open boundary conditions. However, due to quantum interference phenomena, these noise induced links enhance/suppress the transfer process through the chain with periodic boundary conditions depending on its even/odd number of qubits.
\par
The paper is organized as follows. We introduce the model and the master equation governing its dynamics in Sec. \ref{sec:model}; then we describe the strategy to solve such equation and consider the fidelity as a measure of information transfer efficiency. In Sec. \ref{sec:OBC}, we present the solution for XY chain with open boundary conditions, analytical for the smallest non trivial length (3 qubits) and numerical for longer (up to 10 qubits). The results for chains with periodic boundary conditions are provided in Sec.\ref{sec:CBC}, 
again analytical for the smallest non trivial length (3 qubits) and numerical for longer (up to 10 qubits). Finally, conclusions are drawn in Sec. \ref{Conclusion}.

\section{The model} \label{sec:model}
We shall consider a chain of qubits with nearest-neighbor Heisenberg-XY interactions, a model realized in both condensed-matter [\onlinecite{ref:IABVLS}]  
and quantum computing [\onlinecite{ref:ZG}].
The Hamiltonian of the Heisenberg-XY chain is given by
\begin{equation}\label{xyHami}
H_{xy}=\xi\sum_{n=1}^{N^{\prime}}\big(\sigma_{n}\sigma_{n+1}^{\dag}+\sigma_{n}^{\dag}\sigma_{n+1}\big),  
\end{equation}
where $\sigma_{i}:=\ket{0}\bra{1}$ and $\ket{0}$ (resp. $\ket{1}$) is the ground (resp. excited) state of the $i$th qubit and $\xi$ represents the coupling strength between nearest-neighbor qubits. Furthermore $N^{\prime}$ for chains with open and periodic boundary condition equals $N-1$ and $N$ respectively, being $N$ the total number of qubits in (length of) the chain (in closed boundary conditions the $(N+1)$th qubit coincides with the $1$st). 
The Hamiltonian \eqref{xyHami} is hence defined on the Hilbert space $\cal H$ (of $N$ qubits)
 spanned by  $\otimes_{i=1}^{N}\{\ket{0}_{i},\ket{1}_{i}\}$.
\par
Realistic quantum systems are generally open, always exposed to surrounding environment. Thus, we must take into account the influence of external environment to get a proper understanding of the real dynamics of quantum physical systems. 
Here we shall consider quasi-local (or chained) environments affecting the Heisenberg-XY chain as well as local environments.

\begin{figure}[H]
\includegraphics[trim=4cm 15cm 2cm 6cm,clip=true,width=9cm]{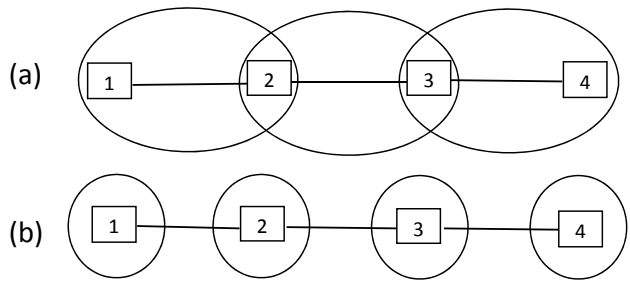}
\caption{
Schematic diagram of 4-qubit chain with chained environments (a) and local environments (b). Rectangles denote qubits, straight lines XY interaction, and  
ellipses/circles environments.}
\label{fig:chain}
\end{figure}

In the former case, as shown in Fig.\ref{fig:chain}a, each individual qubit dissipates into two environments: one common with its left and another with its right nearest neighbor [\onlinecite{ref:LS2011}]. 
The time evolution of the global system consisting of $N$ qubits that interact with each other according to Eq.\eqref{xyHami} as well as with chained environment is described by the following master equation:
\begin{eqnarray}\label{D1}
\dot\rho=\mathcal{D}_{1}[\rho]&=&-i\left[H_{xy},\rho\right]\notag\\
&+&\sum_{n=1}^{N^{\prime}}\gamma\left[
2\left(\sigma_n+\sigma_{n+1}\right)\rho \left(\sigma_n^\dag+\sigma_{n+1}^\dag\right)\right.\notag\\
&-&\left.\left\{\left(\sigma_n^\dag+\sigma_{n+1}^\dag\right)\left(\sigma_n+\sigma_{n+1}\right),\rho\right\}\right],
\end{eqnarray}
in which $\gamma$ is the dissipative parameter and $\{\,,\,\}$ denotes the anti-commutator.
\par
In the latter case, as can be seen in Fig.\ref{fig:chain}b, each individual qubit dissipates only, and independently, into its local environment. The master equation governing the system dynamics of the Heisenberg-XY chain in presence of local environments can be written as
\begin{eqnarray}\label{D2}
\dot\rho=\mathcal{D}_{2}[\rho]&=&-i\left[H_{xy},\rho\right]\notag\\
&+&\sum_{n=1}^{N}\gamma\left[
2\sigma_n\rho \sigma_n^\dag
-\left\{\sigma_n^\dag\sigma_n,\rho
\right\}
\right].
\end{eqnarray}
In order to solve Eqs. \eqref{D1} and \eqref{D2}, we follow the same approach used in  [\onlinecite{ref:LS2013}]. In the sense that we start with the formal solutions $\rho(t)=e^{t\mathcal{D}_{i}}\rho(0)$, $i=1,2$ and then substitute Taylor expansion of $e^{t\mathcal{D}_{i}}$:
\begin{equation}\label{rhot}
\rho(t)=\rho(0)+t\mathcal{D}_{i}\rho(0)+\frac{t^2}{2!}\mathcal{D}_{i}^{2}\rho(0)+\frac{t^3}{3!}\mathcal{D}_{i}^{3}\rho(0)+...
\end{equation}
As can be seen, repeated applications of the super operator ${\cal D}_{i}$ to the initial state $\rho(0)$ will result in the state within the subspace of $\mathbb{H}_{\rho(0)}\subset\mathbb{H}$ where $\mathbb{H}=\mathcal{H}\otimes\mathcal{H}^{*}$  (being $\mathcal{H}^{*}$ the dual of 
$\mathcal{H}$).
In the case of small number of initial excitations $e$ ($e<<N$) that implies $dim(\mathbb{H}_{\rho_{0}})<<dim(\mathbb{H})$ [\onlinecite{ref:LS2014}], 
we can expect to find analytical solutions.
Therefore, applying the super operator $\mathcal{D}_{i}$ to the initial state $\rho(0)$, we achieve closed relations determining a complete set of operators $\{\Pi_i\}$ spanning $\mathbb{H}_{\rho(0)}$, that is $\mathbb{H}_{\rho(0)}=Span\{\Pi_i\}$. Hence $\rho(t)$ can be expanded in terms of them: 
\begin{equation}\label{bases}
\rho(t)=\sum_i a_i(t)\Pi_i.
\end{equation}
Substituting this into the corresponding master equation (Eq. \eqref{D1} or Eq. \eqref{D2}), we find a set of coupled ordinary differential equations for the coefficients $a_{i}(t)$.

Once we have the density operator at any time, we consider the optimal average fidelity between input and output states as measuring the goodness of information transfer.
We set the initial state of the chain as
\begin{eqnarray}
\rho(0)=\ket{\psi}_1\bra{\psi}\otimes \ket{0}_2\bra{0}\otimes \ldots\otimes \ket{0}_N\bra{0},
\end{eqnarray}
where
\begin{eqnarray}\label{initialstate}
\ket{\psi}=\cos(\theta/2)\ket{0}+\sin(\theta/2)e^{i\phi}\ket{1}.
\end{eqnarray}
Qubit 1 is considered as the input, while we label by $o$
the output. It will be $o = N$ for open boundary conditions and 
$o = \lceil\frac{N}{2}\rceil + 1$ for closed boundary conditions.
In such a way the output qubit will always be the farthest
from the input one.

After time $t$, the output state is obtained by reduced density operator of the output qubit $\rho_{o}(t)={\rm Tr}_{\not o}\,\rho(t)$, where ${\rm Tr}_{\not o}$ means the trace overall qubit of the chain but the output ($o$th) one.
 As we will show the dynamics imposes the following general form for output state at qubit $o$
in the basis $\{|0\rangle,|1\rangle\}$:
\begin{equation}\label{rho-out}
 \rho_{o}(t) = \left( \begin{matrix} \varrho(t)\sin^2(\theta/2) &\varsigma(t)\sin\theta e^{i\phi}\\ 
\varsigma(t)^*\sin\theta e^{-i\phi}& 1-\varrho(t)\sin^2(\theta/2) 
\end{matrix} \right).
\end{equation}
Furthermore, we allow the possibility of acting a suitable unitary transformation $V$ on the output qubit. 
Therefore, the input-output fidelity is obtained as [\onlinecite{ref:RJ}]
\begin{equation}\label{fidelity}
f_{1,o}^{V}=\bra{\psi}V\rho_{o}(t)V^{\dag}\ket{\psi},
\end{equation}
where $V$ takes the general form $V = \left( \begin{matrix} u^{*}&-v^{*}\\ v&u \end{matrix} \right)$ with $|u|^{2}+|v|^{2}=1$. 
As we will see later $V$ allows us to introduce phase shifts on the off-diagonal elements of the output density matrix and affects the fidelity [\onlinecite{ref:MLM}].
The average fidelity can be found by integrating over all input states, that is over the Bloch sphere:
\begin{align}\label{averagefidelity}
 &F_{1,o}^{V}=\frac{1}{4\pi}\int_{0}^{\pi}d\theta \sin(\theta)\int_{0}^{2\pi}d\phi f_{1,o}^{V}(\theta,\phi,t)&&\nonumber\\
&\quad\quad=\frac{1}{2}+\frac{1}{6}\left\{2\varrho|u|^2+2(\varsigma u^{*2}+\varsigma^{*}u^{2})-\varrho\right\}.
\end{align}
We then maximize it over $V$ obtaining the optimal value of $u$ to be
\begin{eqnarray}\label{Vtransform}
u_{opt}=e^{i\left\{1/2\tan^{-1}[Im(\varsigma)/Re(\varsigma)]+{\rm sgn}[Re(\varsigma)]\pi/2\right\}}.
\end{eqnarray}
Hence the optimal average fidelity $F_{1,o}^{opt}$ reads as 
\begin{align}\label{F13opt}
F_{1,o}^{opt}=\frac{1}{2}+\frac{1}{6}\left\{\varrho+4\sqrt{[Re(\varsigma)]^2+[Im(\varsigma)]^2}\right\}.
\end{align}

\section{Open boundary conditions}\label{sec:OBC}
In this section, we shall consider a Heisenberg-XY chain with open boundary condition in presence of quasi-local and local environments. First, we shall analytically calculate the optimal average fidelity for the shortest non trivial chain with 3 qubits. Then, we shall provide the results for longer chains by numerical calculations and analyze the effects of the local and chained environments on the state transfer process.
\par
\subsection{Quasi-local environments}
The master equation \eqref{D1} for a 3-qubit chain with open boundary condition reads
\begin{align}\label{OBCchained}
&\dot\rho=\mathcal{D}_{1}[\rho]=-i\xi\left[\left(\sigma_1\sigma_2^\dag + \sigma_1^\dag\sigma_2\right)
+\left(\sigma_2\sigma_3^\dag + \sigma_2^\dag\sigma_3\right),\rho\right]\notag\\
&+\gamma\left[
2\left(\sigma_1+\sigma_2\right)\rho \left(\sigma_1^\dag+\sigma_2^\dag\right)
-\left\{\left(\sigma_1^\dag+\sigma_2^\dag\right)\left(\sigma_1+\sigma_2\right),\rho\right\}
\right]\notag\\
&+\gamma\left[
2\left(\sigma_2+\sigma_3\right)\rho \left(\sigma_2^\dag+\sigma_3^\dag\right)
-\left\{\left(\sigma_2^\dag+\sigma_3^\dag\right)\left(\sigma_2+\sigma_3\right),\rho\right\}
\right].\notag\\
\end{align}
Using the strategy mentioned in Sec.\ref{sec:model}, we get the set of operators $\{\Pi_{i}\}$ that span $\mathbb{H}_{\rho(0)}$ as follows:
\begin{align}
	&\Pi_0=\ket{0}\bra{0},\notag\\
	&\Pi_k=\ket{k}\bra{k},\notag\\
        &\Pi_{2k+2}=\ket{k}\bra{0}+\ket{0}\bra{k},\notag\\
         &\Pi_{2k+3}=i(\ket{k}\bra{0}-\ket{0}\bra{k}),\notag\\
         &\Pi_{2k+2l+4}=\ket{k}\bra{l}+\ket{l}\bra{k},\notag\\
         &\Pi_{2k+2l+5}=i(\ket{k}\bra{l}-\ket{l}\bra{k}),
         \label{basesopenchain}
\end{align}
with $k,l=1,2,3$ and $l<k$. Here $\ket{k}$ stands for the chain state with a single excitation located on the $k$th qubit. 
Expanding the density matrix as Eq.\eqref{bases} and inserting into Eq.\eqref{OBCchained}, we find a set of coupled ordinary differential equations for the coefficients that are reported in Appendix A. Just coefficients 
$a_{3},a_{8},a_{9}$ appear in output state. Solving the differential equations for these coefficients (see appendix A) 
we find the output state \eqref{rho-out} with
\begin{align}\label{rho1OBCchained}
&\varrho(t)=\frac{a_{3}}{\sin^{2}(\theta/2)}&&\nonumber\\
&\hspace{0.5cm}=\frac{1}{4}\left\{e^{-2\gamma t}-2e^{-5\gamma t/2}\Big[\cosh(\frac{\gamma t x}{2})\cos(\frac{\gamma ty}{2})\right.&&\nonumber\\
&\hspace{0.5cm}+\left.\frac{y\cosh(\frac{\gamma t x}{2})\sin(\frac{\gamma ty}{2})+x\cos(\frac{\gamma ty}{2})\sinh(\frac{\gamma tx}{2})}{x^2+y^2}\Big]\right.&&\nonumber\\
&\hspace{0.5cm}+\left.\frac{1}{2} e^{-3\gamma t}\Big[\cos(\gamma ty)+\cosh(\gamma tx)\right.&&\nonumber\\
&\hspace{0.5cm}+\left.\frac{2y\sin(\gamma ty)+2x\sinh(\gamma tx)-\cos(\gamma ty)+\cosh(\gamma tx)}{x^2+y^2}\Big]\right\},&&
 \end{align}
and
\begin{align}\label{rho2OBCchained}
&\varsigma(t)=\frac{a_{8}+ia_{9}}{\sin(\theta)e^{i\phi}}=\frac{-e^{-\gamma t}}{4}&&\nonumber\\
&\hspace{0.5cm}+\frac{e^{-3\gamma t/2}}{4}\cosh\Big(\frac{\gamma t}{2}(x+iy)\Big)\Big[1
+\frac{\tanh\Big(\frac{\gamma t}{2}(x+iy)\Big)}{x+iy}\Big],&&
 \end{align}
where
 \begin{align}\label{xydefine}
&x=\left\{\frac{9-8(\frac{\xi}{\gamma})^2+\sqrt{81+112(\frac{\xi}{\gamma})^2+64(\frac{\xi}{\gamma})^4}}{2}\right\}^{1/2},&&\nonumber\\
&y=\left\{\frac{-9+8(\frac{\xi}{\gamma})^2+\sqrt{81+112(\frac{\xi}{\gamma})^2+64(\frac{\xi}{\gamma})^4}}{2}\right\}^{1/2}.&&
\end{align}
Inserting Eqs.\eqref{rho1OBCchained} and \eqref{rho2OBCchained} into \eqref{F13opt}, we get
\begin{eqnarray}\label{FOBCchain}
F_{1,3}^{opt}&=&\frac{1}{2}+\frac{1}{24}\Big[\left\{e^{-2\gamma t}+\frac{1}{2}e^{-3\gamma t}[\cos(\gamma t y)+\cosh(\gamma t x)]\right.\cr\cr
&-&\left. 2e^{-5\gamma t/2}\cosh(\frac{\gamma t x}{2})\cos(\frac{\gamma t y}{2})+\frac{e^{-3\gamma t}}{x^2+y^2}\right.\cr\cr
&\times&\left.\Big(\frac{1}{2}[\cosh(\gamma t x)-\cos(\gamma t y)]+x \sinh(\frac{\gamma t x}{2})\right.\cr\cr
&+&\left.y \sin(\frac{\gamma t y}{2})-2e^{\gamma t/2}[x\sinh(\frac{\gamma t x}{2})\cos(\frac{\gamma t y}{2})\right.\cr\cr
&+&\left.y\cosh(\frac{\gamma t x}{2})\sin(\frac{\gamma t y}{2})]\Big)\right\}^{1/2}+1\Big]^2-\frac{1}{24}.
\end{eqnarray}

\subsection{Local environments}

We rewrite Eq.\eqref{D2} for a chain of three qubits with open boundary condition 
\begin{align}\label{OBClocal}
&\dot\rho=\mathcal{D}_{2}[\rho]=-i\xi\left[\left(\sigma_1\sigma_2^\dag + \sigma_1^\dag\sigma_2\right)
+\left(\sigma_2\sigma_3^\dag + \sigma_2^\dag\sigma_3\right),\rho\right]\notag\\
&+\gamma\left[2\sigma_{1}\rho\sigma_{1}^{\dag}-\sigma_{1}^{\dag}\sigma_{1}\rho-\rho\sigma_{1}^{\dag}\sigma_{1}
\right]\notag\\
&+\gamma\left[2\sigma_{2}\rho\sigma_{2}^{\dag}-\sigma_{2}^{\dag}\sigma_{2}\rho-\rho\sigma_{2}^{\dag}\sigma_{2}
\right]\notag\\
&+\gamma\left[2\sigma_{3}\rho\sigma_{3}^{\dag}-\sigma_{3}^{\dag}\sigma_{3}\rho-\rho\sigma_{3}^{\dag}\sigma_{3}
\right].
\end{align}
Then, the action of $\mathcal{D}_{2}$ on the initial state $\rho(0)$ of three qubits (of which the first is the input one)  leads to the subspace $\mathbb{H}_{\rho(0)}$ spanned by
\begin{align}
	&\Pi_0=\ket{0}\bra{0},\notag\\
	&\Pi_k=\ket{k}\bra{k},\notag\\
        &\Pi_{2k+2}=\ket{k}\bra{0}+\ket{0}\bra{k},\notag\\
         &\Pi_{2k+3}=i(\ket{k}\bra{0}-\ket{0}\bra{k}),\notag\\
         &\Pi_{10}=i(\ket{2}\bra{1}-\ket{1}\bra{2}),\notag\\
         &\Pi_{11}=\ket{3}\bra{1}+\ket{1}\bra{3},\notag\\
         &\Pi_{12}=i(\ket{3}\bra{2}-\ket{2}\bra{3}),
         \label{basesopenlocal}
\end{align}
with $k=1,2,3$.
Substituting Eq.\eqref{bases} into Eq.\eqref{OBClocal} we get the set of ordinary differential equations reported in Appendix A together with its solutions.
Using them we arrive at the elements of the reduced density operator for the third (output)
qubit
 \begin{eqnarray}\label{rho12OBClocal}
\varrho &=&e^{-2\gamma t}\sin^4\left(\frac{\xi t}{\sqrt{2}}\right),\cr\cr
\varsigma&=&-\frac{1}{2}e^{-\gamma t}\sin^2\left(\frac{\xi t}{\sqrt{2}}\right).\notag\\
 \end{eqnarray}
Finally, thanks to Eg.\eqref{F13opt}, the optimal average fidelity reads:
\begin{align}\label{Fopenlocal}
&F_{1,3}^{opt}=\frac{1}{2}+\frac{1}{6}\left\{\varrho+4\abs{\varsigma}\right\}&&\nonumber\\
&\hspace{0.7cm}=\frac{1}{2}+\frac{e^{-\gamma t}}{3}\sin^{2}\left(\frac{\xi t}{\sqrt{2}}\right)\left[1+
\frac{e^{-\gamma t}}{2}\sin^{2}\left(\frac{\xi t}{\sqrt{2}}\right)\right].
\end{align}
\par
\begin{figure}[H]
\centering
\subfloat[$\xi=1$,$\gamma=4$]{
\includegraphics[width=6cm]{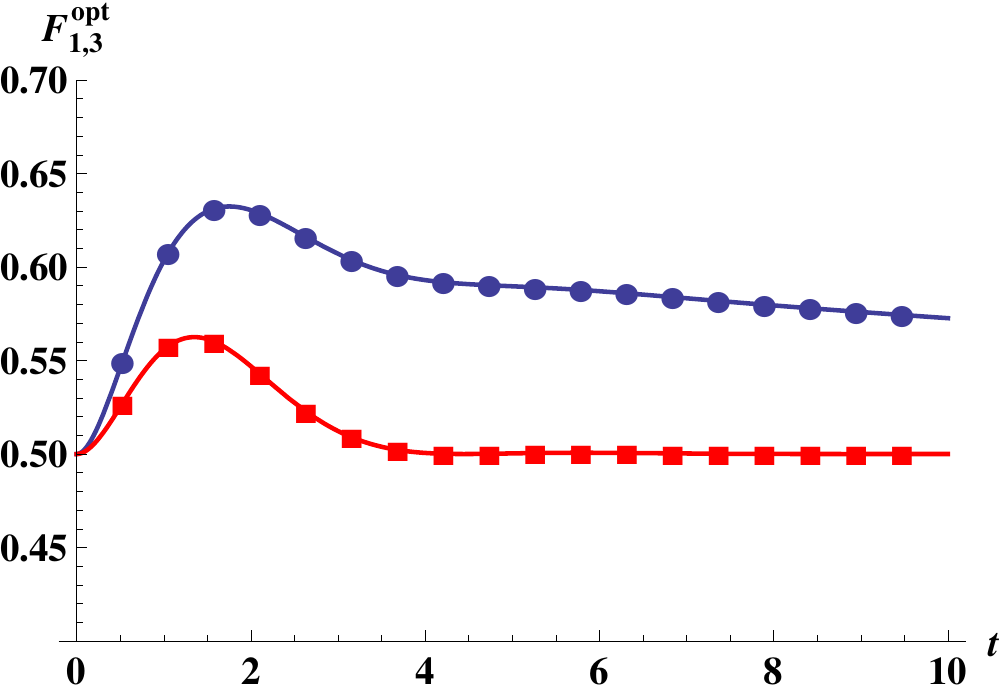}}
\qquad
\subfloat[$\xi=1$,$\gamma=20$]{
\includegraphics[width=6cm]{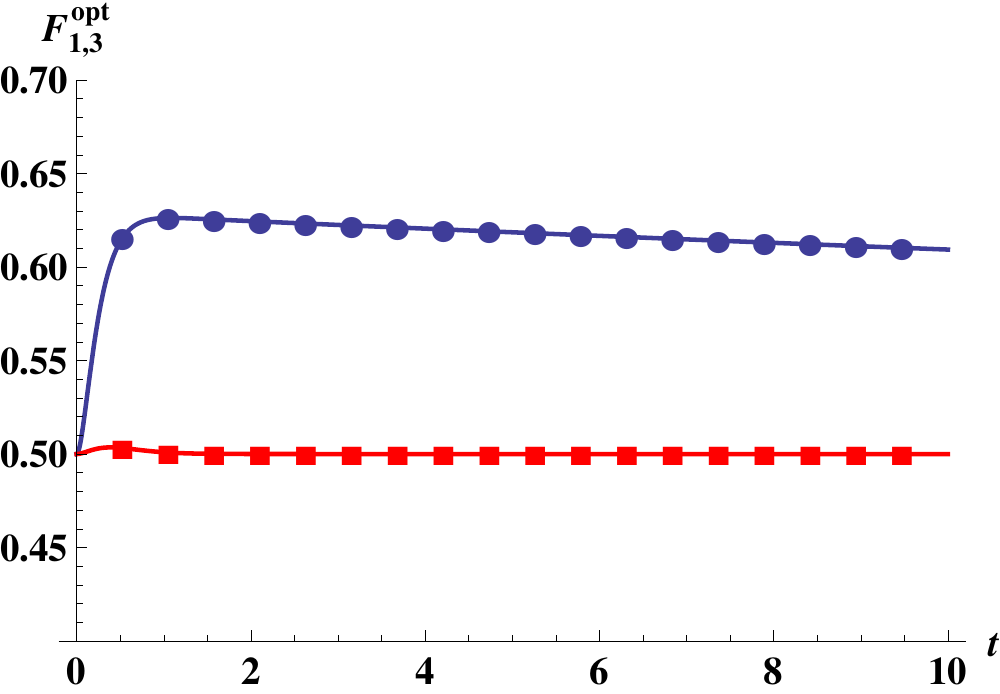}}
\caption{\label{fig:threeOBC}Optimal average fidelity between first and third qubits of a three-qubit Heisenberg-XY chain with open boundary conditions in presence of chained (blue dots) and local (red squares) noise.}
\centering
\end{figure}

We are now going to compare the effects of local and quasi-local noise on  information transfer in Heisenberg-XY chain. Fig.\ref{fig:threeOBC} (a,b) show Eq.\eqref{FOBCchain} (blue dots) and 
Eq.\eqref{Fopenlocal} (red squares) in two different noise regimes.
\par
 According to Fig.\ref{fig:threeOBC} (a), when $\gamma$ is comparable to $\xi$ we observe residual oscillatory behavior that comes from the Heisenberg-XY Hamiltonian dynamics. 
 Also, chained environments induce indirect links between nearest-neighbor qubits and increase the fidelity with respect to the local environments. 
The behavior in strong-noise regime ($\gamma>>\xi$), Fig.\ref{fig:threeOBC} (b), for local environments can be interpreted by the quantum Zeno effect. That is, repeated measurements 
on the quantum system can freeze its Hamiltonian evolution [\onlinecite{ref:FP}]. Here the strong interaction of the quantum chain with its local environments, playing the role of measuring apparatus, effectively decouples each qubit from its nearest neighbors in the chain and leads to the lowest optimal average fidelity. In contrast, chained environments inducing long-lived indirect interactions between contiguous qubits significantly enhance the optimal average fidelity and quantum state transfer process.
\par
\begin{figure}[H]
\centering
\subfloat[$\xi=1$,$\gamma=4$]{
\includegraphics[width=6cm]{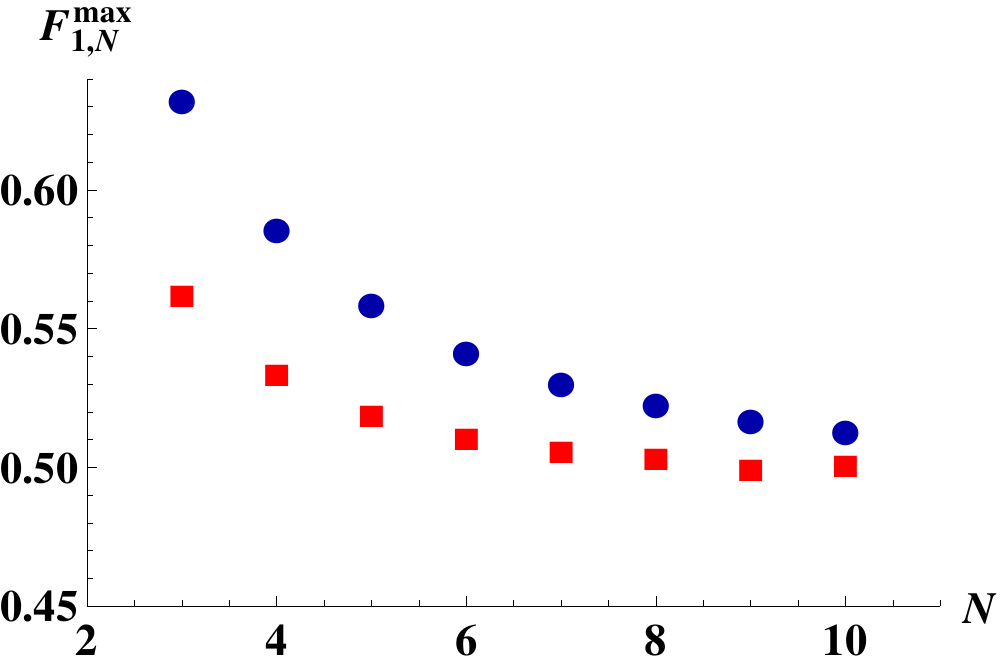}}
\qquad
\subfloat[$\xi=1$,$\gamma=20$]{
\includegraphics[trim=0cm 0cm 0cm 0cm,clip=true,width=6cm]{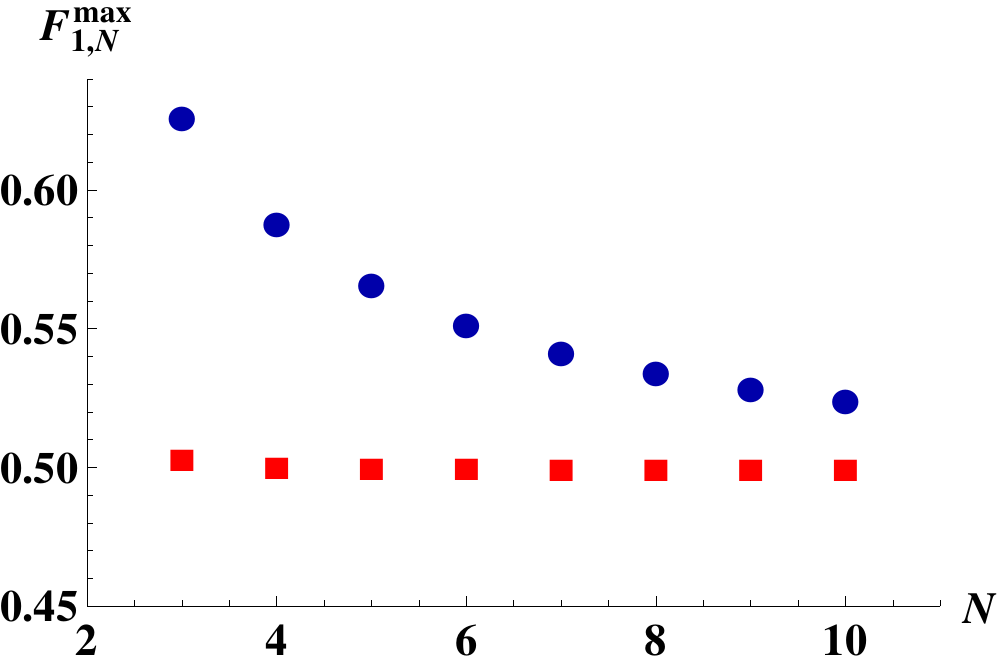}}
\caption{\label{fig:OBC}Maximum of optimal average fidelity between the two ends of the chain vs number $N$ of qubits with open boundary conditions in presence of quasi-local (blue dots) and local (red squares) envornments.}
\centering
\end{figure}

 
Next, we have numerically solved the differential equations in $\mathbb{H}_{\rho(0)}$ obtained by Eqs.\eqref{D1} and \eqref{D2} for system's size $N=4,...,10$, found the optimal average fidelity between the first and last qubits of the chain and represented its maximal value for different chain lengths in Fig.\ref{fig:OBC}.
\par
According to Fig.\ref{fig:OBC}, the fidelity of state transfer in both cases of local and quasi-local noise decreases by increasing the chain's length. This behavior, that is also reported in [\onlinecite{ref:BS}] for Heisenberg-XY chain in the absence of noise, comes from the dispersion of information over the chain (see also [\onlinecite{ref:BS05}]). 
The better performance of quasi-local environment with respect to the local one (which is more evident  in the strong noise regime) tends to be washed out over a length of more than $10$ qubits.


\section{Periodic boundary condition} \label{sec:CBC}
In this section, we study the effects of local and quasi-local environments on the efficiency of state transfer through a chain having closed boundary conditions.
Likewise the previous section, analytical calculations for the fidelity in the case of three qubits will be provided and then the fidelity for longer chains will be evaluated numerically.

\subsection{Quasi-local environments}
Eq.\eqref{D1} for a three-qubit chain with closed boundary condition becomes: 
\begin{align}\label{CBCchained}
&\dot\rho=\mathcal{D}_{1}[\rho]=-i\xi\left[\left(\sigma_1\sigma_2^\dag + \sigma_1^\dag\sigma_2\right)
+\left(\sigma_2\sigma_3^\dag + \sigma_2^\dag\sigma_3\right)\right.\notag\\
&\hspace{3cm}\left.+\left(\sigma_3\sigma_1^\dag + \sigma_3^\dag\sigma_1\right),\rho\right]\notag\\
&+\gamma\left[
2\left(\sigma_1+\sigma_2\right)\rho \left(\sigma_1^\dag+\sigma_2^\dag\right)
-\left\{\left(\sigma_1^\dag+\sigma_2^\dag\right)\left(\sigma_1+\sigma_2\right),\rho\right\}
\right]\notag\\
&+\gamma\left[
2\left(\sigma_2+\sigma_3\right)\rho \left(\sigma_2^\dag+\sigma_3^\dag\right)
-\left\{\left(\sigma_2^\dag+\sigma_3^\dag\right)\left(\sigma_2+\sigma_3\right),\rho\right\}
\right]\notag\\
&+\gamma\left[
2\left(\sigma_3+\sigma_1\right)\rho \left(\sigma_3^\dag+\sigma_1^\dag\right)
-\left\{\left(\sigma_3^\dag+\sigma_1^\dag\right)\left(\sigma_3+\sigma_1\right),\rho\right\}
\right].\notag\\
\end{align}
Applying the operator $\mathcal{D}_{1}$ on $\rho(0)$ yields the complete set of operators as in Eq.\eqref{basesopenchain} and a set of ordinary differential equations that are reported in Appendix \ref{appB} together with its solutions.
Then the reduced density operator for the third qubit results
as Eq.\eqref{rho-out} with:
\begin{eqnarray}\label{rhoCBCchained}
\varrho &=&\frac{e^{-2\gamma t}}{9}\left\{1+e^{-6\gamma t}-2e^{-3t\gamma}\cos(3t\xi)\right\},\cr\cr
\varsigma &=&\frac{e^{2\gamma t(i\xi-\gamma)}}{6}\left\{e^{-6\gamma t(i\xi+\gamma)}-1\right\}.
 \end{eqnarray}

Finally, using Eq.\eqref{F13opt}, the optimal average fidelity can be found as
\begin{eqnarray}\label{FCBCchained}
F_{1,3}^{opt}&=&\frac{1}{54}\left[\frac{1}{3}\sqrt{e^{-2\gamma t}+e^{-8\gamma t}-2e^{-5\gamma t}\cos(3t\xi)}+1\right]^2\notag\\
&-&\frac{1}{54}+\frac{1}{2}.
\end{eqnarray}

\subsection{Local environments}
The master equation governing the dynamics of a chain of three-qubit with periodic boundary conditions is obtained from Eq.\eqref{D2} as
\begin{align}\label{CBClocal}
&\dot\rho=\mathcal{D}_{2}[\rho]=-i\xi\left[\left(\sigma_1\sigma_2^\dag + \sigma_1^\dag\sigma_2\right)
+\left(\sigma_2\sigma_3^\dag + \sigma_2^\dag\sigma_3\right)\right.\notag\\
&\hspace{1.5cm}\left.+\left(\sigma_3\sigma_1^\dag + \sigma_3^\dag\sigma_1\right),\rho\right]\notag\\
&\hspace{1.5cm}+\gamma\left[2\sigma_{1}\rho\sigma_{1}^{\dag}-\sigma_{1}^{\dag}\sigma_{1}\rho-\rho\sigma_{1}^{\dag}\sigma_{1}
\right]\notag\\
&\hspace{1.5cm}+\gamma\left[2\sigma_{2}\rho\sigma_{2}^{\dag}-\sigma_{2}^{\dag}\sigma_{2}\rho-\rho\sigma_{2}^{\dag}\sigma_{2}
\right]\notag\\
&\hspace{1.5cm}+\gamma\left[2\sigma_{3}\rho\sigma_{3}^{\dag}-\sigma_{3}^{\dag}\sigma_{3}\rho-\rho\sigma_{3}^{\dag}\sigma_{3}
\right].\notag\\
\end{align}
Then, one can arrive at the Eq.\eqref{basesopenchain} and 
a set of ordinary differential equations 
for coefficients $a_i$ that are reported in Appendix B together with its solutions. We can then find the elements of the reduced density operator of the output (third) qubit:
\begin{eqnarray}\label{rhoCBClocal}
\varrho&=&\frac{4}{9}e^{-2\gamma t}\sin^{2}(3t\xi/2),\cr\cr
\varsigma&=&-\frac{i}{3}e^{-\gamma t-it\xi/2}\sin(3t\xi/2).
 \end{eqnarray}
The optimal average fidelity results
\begin{eqnarray}\label{FlocalCBC}
F_{1,3}^{opt}&=&\frac{1}{2}-\frac{1}{3}e^{-2\gamma t}\left\{\frac{1}{2}-\frac{2}{3}\sin^{2}\left(\frac{3t\xi}{2}\right)\right.\cr\cr
&-&\left.\frac{1}{18}[5+4\cos\left(3t\xi\right)]-\frac{2}{3}e^{\gamma t}\left|\sin\left(\frac{3t\xi}{2}\right)\right|\right\}.\notag\\
\end{eqnarray}

\begin{figure}[H]
\centering
\subfloat[$\xi=1$,$\gamma=4$]{
\includegraphics[width=6cm]{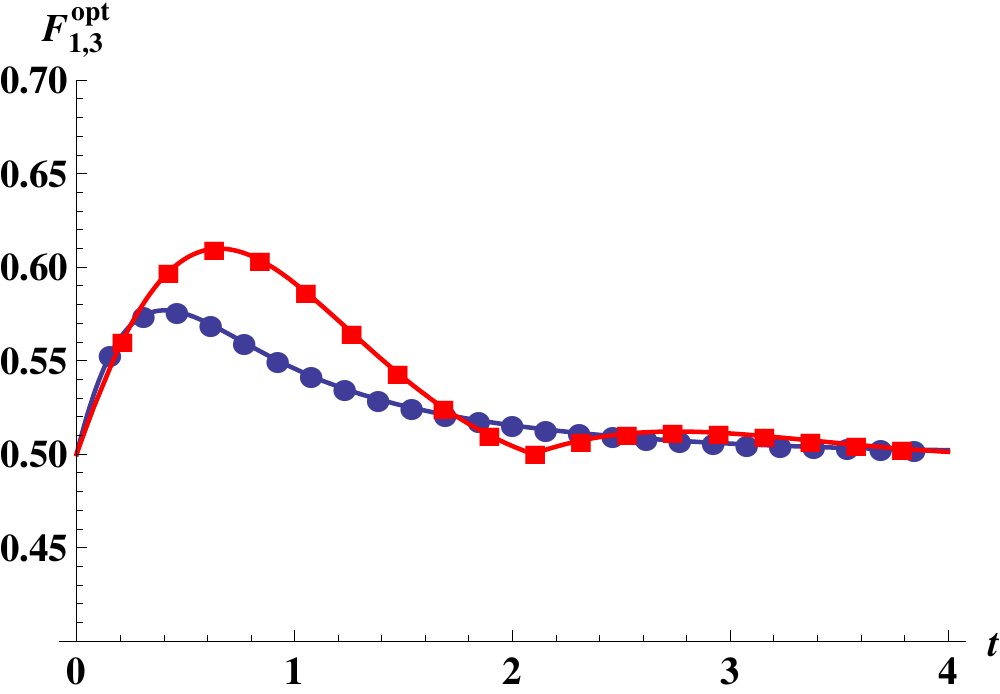}}
\qquad
\subfloat[$\xi=1$,$\gamma=20$]{
\includegraphics[width=6cm]{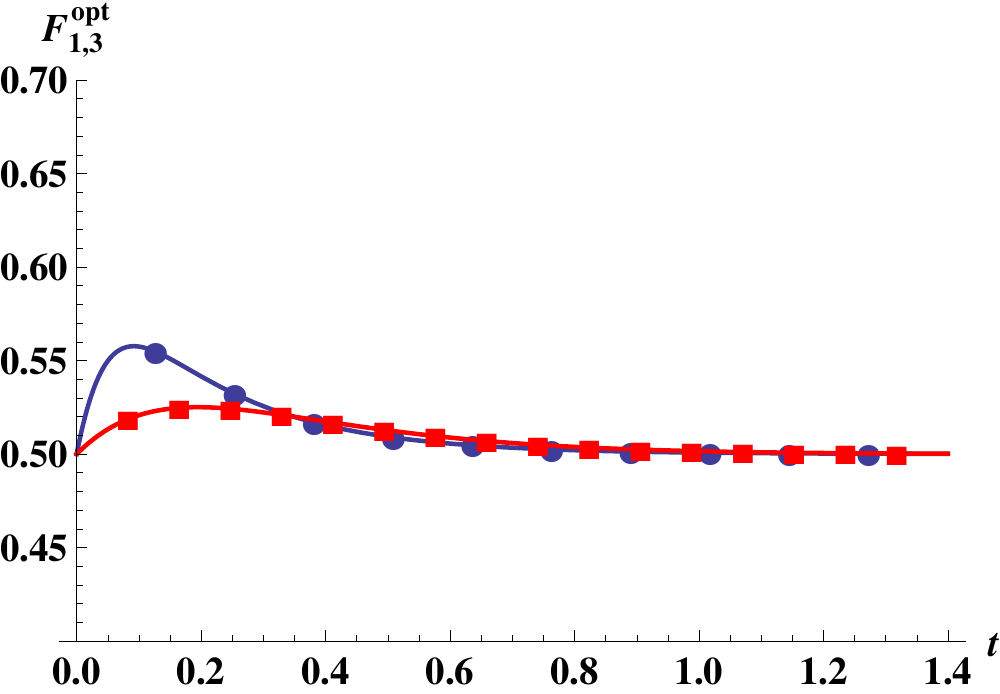}}
\caption{\label{fig:3CBC}Optimal average fidelity between the first and third qubits of a closed three-qubit chain with local (red squares) and chained (blue dots) environments.
}
\centering
\end{figure}

At the end, we compare the influence of local and quasi-local noise on the optimal average fidelity.
In Fig.\ref{fig:3CBC}, we report the optimal average fidelity \eqref{FCBCchained} and \eqref{FlocalCBC} for two different noise regimes.
\par
\begin{figure}[H]
\centering
\subfloat[$\xi=1$,$\gamma=4$]{
\includegraphics[width=6cm]{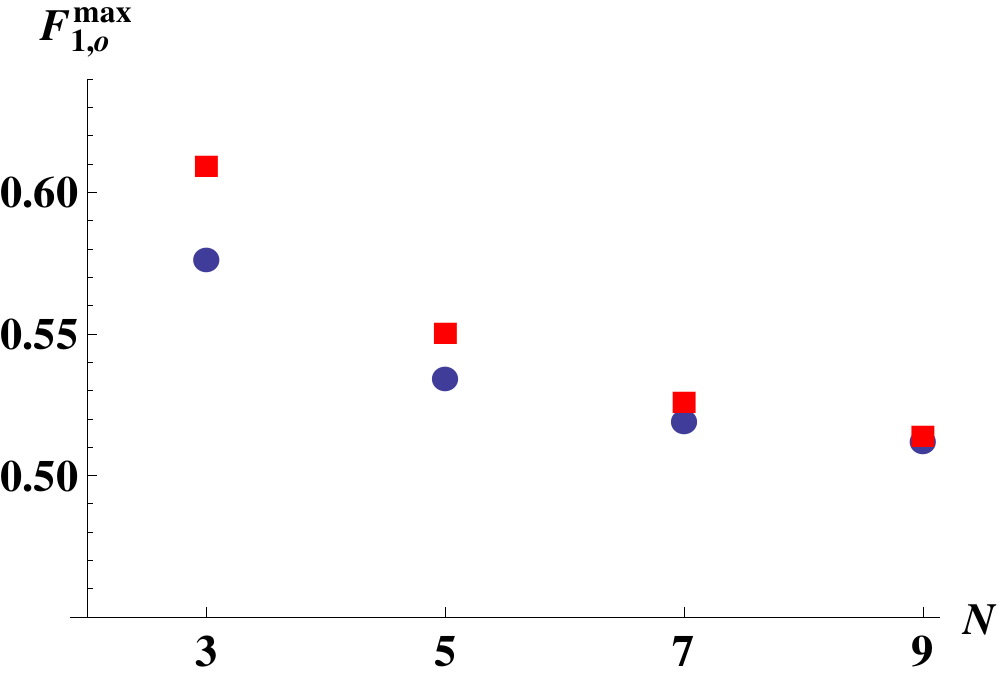}}
\qquad
\subfloat[$\xi=1$,$\gamma=20$]{
\includegraphics[width=6cm]{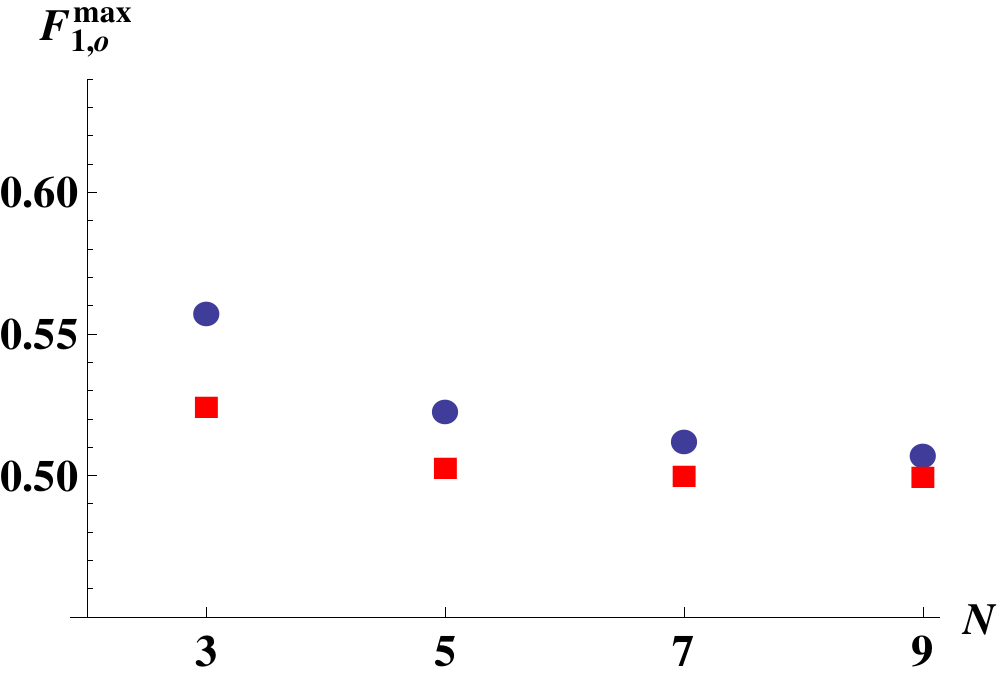}}
\caption{\label{fig:CBCodd}
Maximum of optimal average fidelity between the first qubit and its farthest neighbor in a closed chain of odd size ($N=3,5,7,9$). Here $o=\lceil\frac{N}{2}\rceil+1$  labels the output qubit affected by chained (blue dots) and local (red squares) environments.}
\end{figure}
\begin{figure}[H]
\centering
\subfloat[$\xi=1$,$\gamma=4$]{
\includegraphics[width=6.25cm]{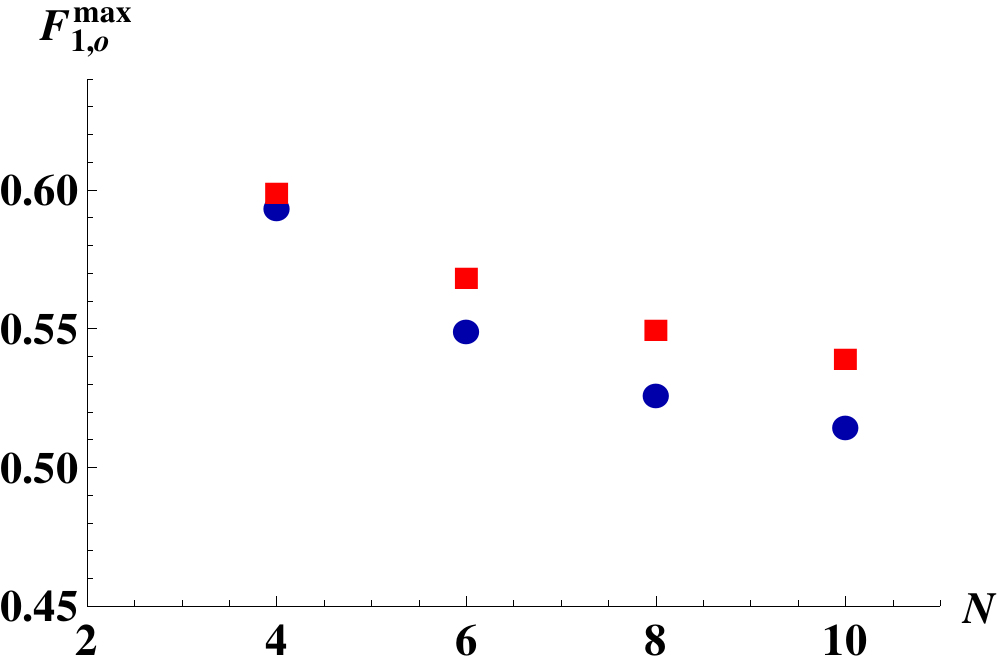}}
\qquad
\subfloat[$\xi=1$,$\gamma=20$]{
\includegraphics[width=6.25cm]{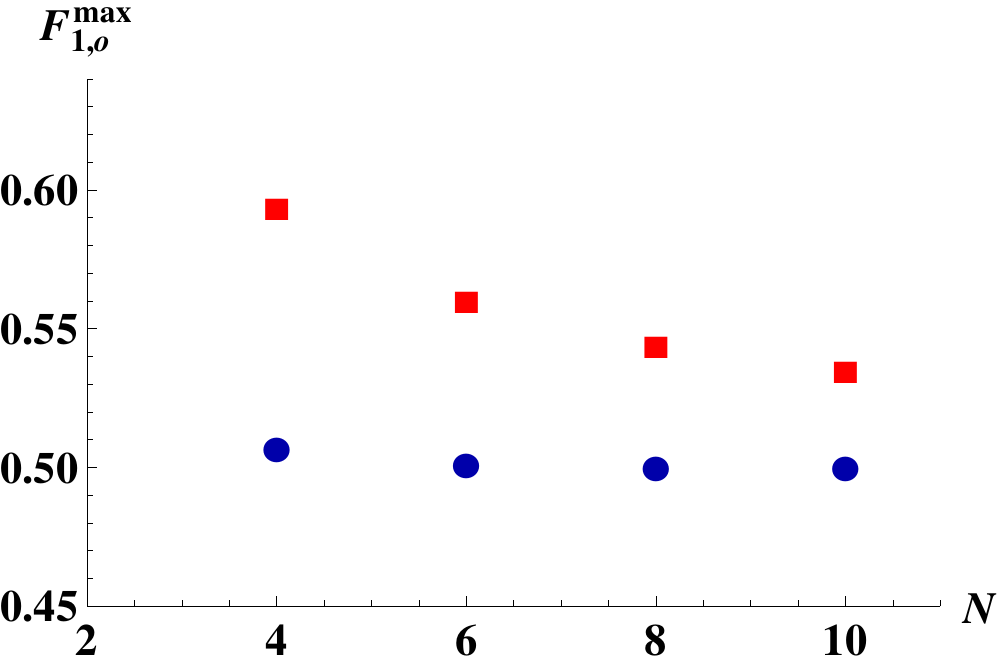}}
\caption{\label{fig:CBCeven}
Maximum of optimal average fidelity between the first qubit and its farthest neighbor in a closed chain of even size ($N=4,6,8,10$). Here $o=\lceil\frac{N}{2}\rceil+1$ labels the output qubit affected by chained (blue dots) and local (red squares) environments.}
\end{figure}

As shown in Fig.\ref{fig:3CBC}, when the dissipative parameter is not large enough to establish strong environment-induced chain links in quasi-local case, due to the interference phenomena, local environments are more efficient. To have a clear picture of the effect of these two kind of noise we report the maximum of optimal average fidelity between the $1$st and its farthest qubit for chains with periodic boundary condition of odd and even sizes in Figs.\ref{fig:CBCodd},\ref{fig:CBCeven}. 

In the regime in which $\gamma$ is comparable with $\xi$, both Hamiltonian and dissipative terms play relevant roles in the system's dynamics. The wave function propagating in a closed chain depending on the odd/even number of qubits experiences opposite phases, as a consequence of the quantum interference phenomena, hence the induced links by dimerized environment reduce/enhance the performance of information transfer (see Figs.(\ref{fig:CBCodd}a,\ref{fig:CBCeven}a)).
In the strong dissipative regime ($\gamma>>\xi$), there is weak direct connection between qubits (due to quantum Zeno kind effect) that implies  $F_{1,o}^{max}\approx 0.5$ in the case of local noise. However, the dimerized case exploits the indirect induced connections and significantly enhances the state transfer process on the chain (Figs.(\ref{fig:CBCodd}b,\ref{fig:CBCeven}b)). However, this higher performance is also decreasing by increasing the system's size.

\section{Conclusion}\label{Conclusion}
We have studied the effect of two different kinds of dissipation, quasi-local and local, on quantum state transfer precess in Heisenberg-XY chain with open and closed boundary conditions. We have shown that the chained environments inducing indirect connections between nearest-neighbor qubits remarkably enhance the fidelity of state transfer in short chains ($N<10$) with open boundary conditions. In the situation of chains with closed boundary conditions the same behavior is found in the strong noise regime. However, due to quantum interference phenomena, distinct behaviors appear in the weak-noise regime:  
chained environments decrease/increase the fidelity of state transfer in chains with odd/even number of qubits.
Going beyond the quasi-local dissipation model investigated here, we might claim that the fidelity enhancement can be related to the spatial extension of non-local environment effects.  
The obtained results highlight the relevance of the topology of environmental actions on a set of Hamiltonian interacting qubits. A subject that deserves attention due the continuing miniaturization of quantum devices.

\acknowledgements
R. Radgohar and L. Memarzadeh would like to thank the University of Camerino for hospitality.

\appendix
\section{Density matrix for open boundary conditions}\label{appA}
Here, we shall investigate the dynamics of density matrix of a three-qubit Heisenberg-XY chain with open boundary condition dissipating in quasi-local environments as well as in local environments. In the former case, we expand the density matrix using the operators in Eq.\eqref{basesopenchain} and insert into Eq.\eqref{OBCchained} and get:
\begin{align}
&\dot{a}_{0}(t)=2\gamma\left\{a_{1}(t)+a_{3}(t)+2[a_{10}(t)+a_{14}(t)+a_{2}(t)]\right\},&&\nonumber\\
&\dot{a}_{1}(t)=-2\gamma[a_{1}(t)+a_{10}(t)]+2\xi a_{11}(t),&&\nonumber\\
&\dot{a}_{2}(t)=-2\gamma[a_{10}(t)+a_{14}(t)+2a_{2}(t)]+2\xi[a_{15}(t)-a_{11}(t)],&&\nonumber\\
&\dot{a}_{3}(t)=-2\gamma[a_{3}(t)+a_{14}(t)]-2\xi a_{15}(t),&&\nonumber\\
&\dot{a}_{4}(t)=-\gamma[a_{4}(t)+a_{6}(t)]+\xi a_{7}(t),&&\nonumber\\
&\dot{a}_{5}(t)=-\gamma[a_{5}(t)+a_{7}(t)]-\xi a_{6}(t),&&\nonumber\\
&\dot{a}_{6}(t)=-\gamma[a_{4}(t)+2a_{6}(t)+a_{8}(t)]+\xi[a_{5}(t)+a_{9}(t)],&&\nonumber\\
&\dot{a}_{7}(t)=-\gamma[a_{5}(t)+2a_{7}(t)+a_{9}(t)]-\xi[a_{4}(t)+\xi a_{8}(t)],&&\nonumber\\
&\dot{a}_{8}(t)=-\gamma[a_{6}(t)+a_{8}(t)]+\xi a_{7}(t),&&\nonumber\\
&\dot{a}_{9}(t)=-\gamma[a_{7}(t)+a_{9}(t)]-\xi a_{6}(t),&&\nonumber\\
&\dot{a}_{10}(t)=-\gamma[a_{1}(t)+a_{2}(t)+3a_{10}(t)+a_{12}(t)]+\xi a_{13}(t),&&\nonumber\\
&\dot{a}_{11}(t)=-\gamma[3a_{11}(t)+a_{13}(t)]-\xi[a_{1}(t)-a_{2}(t)+a_{12}(t)],&&\nonumber\\
&\dot{a}_{12}(t)=-\gamma[a_{10}(t)+2a_{12}(t)+a_{13}(t)]+\xi[a_{11}(t)-a_{15}(t)],&&\nonumber
\end{align}
\begin{align}
&\dot{a}_{13}(t)=-\gamma[a_{11}(t)+2a_{13}(t)-a_{15}(t)]-\xi[a_{10}(t)-a_{14}(t)],&&\nonumber\\
&\dot{a}_{14}(t)=-\gamma[a_{2}(t)+a_{3}(t)+a_{12}(t)+3a_{14}(t)]-\xi a_{13}(t),&&\nonumber\\
&\dot{a}_{15}(t)=-\gamma[a_{13}(t)+3a_{15}(t)]-\xi[a_{2}(t)-a_{3}(t)-a_{12}(t)],&&
\end{align}
with the following initial conditions:
\begin{align}\label{initialconditions}
        &a_{0}(0)=\cos^{2}(\theta/2),&&a_{1}=\sin^{2}(\theta/2),\nonumber\\
        &a_{4}(0)=\frac{1}{2}\sin(\theta)\cos(\phi),&&a_{5}(0)=\frac{1}{2}\sin(\theta)\sin(\phi),\nonumber\\
        &a_{i}(0)=0,\forall i\neq0,1,4,5.
\end{align}

We only report solutions for coefficients $a_{3},a_{8},a_{9}$ that are required to calculate the reduced density matrix for the third qubit.
\begin{eqnarray}
a_{3}&=&\frac{1}{4}\left\{e^{-2\gamma t}-2e^{-5\gamma t/2}\Big[\cosh(\frac{\gamma t x}{2})\cos(\frac{\gamma ty}{2})\right.\cr\cr
&+&\left.\frac{y\cosh(\frac{\gamma t x}{2})\sin(\frac{\gamma ty}{2})+x\cos(\frac{\gamma ty}{2})\sinh(\frac{\gamma tx}{2})}{x^2+y^2}\Big]\right.\cr\cr
&+&\left.\frac{1}{2} e^{-3\gamma t}\Big[\cos(\gamma ty)+\cosh(\gamma tx)\right.\cr\cr
&+&\left.\frac{2y\sin(\gamma ty)+2x\sinh(\gamma tx)-\cos(\gamma ty)+\cosh(\gamma tx)}{x^2+y^2}\Big]\right\}\cr\cr
&\times&\sin^{2}(\theta/2),\cr\cr
a_{8}&=&\frac{1}{8}e^{-\frac{3\gamma t}{2}}\left\{-2e^{\gamma t/2}\cos\phi+e^{-i\phi}\cosh[\frac{\gamma t}{2}(x-\frac{16i\xi}{x})]\Big(1\right.\cr\cr
&+&\left.\frac{\tanh[\frac{\gamma t}{2}(x-\frac{16i\xi}{x})]}{x-16i\xi/x}\Big)+e^{i\phi}\cosh[\frac{\gamma t}{2}(x+\frac{16i\xi}{x})]\Big(1+\right.\cr\cr
&+&\left.\frac{\tanh[\frac{\gamma t}{2}(x+\frac{16i\xi}{x})]}{x+\frac{16i\xi}{x}}\Big)\right\}\times\sin(\theta),\cr\cr
a_{9}&=&\frac{1}{8}e^{-\frac{3\gamma t}{2}}\left\{-2e^{\gamma t/2}\sin\phi+ie^{-i\phi}\cosh[\frac{\gamma t}{2}(x-\frac{16i\xi}{x})]\Big(1+\right.\cr\cr
&+&\left.\frac{\tanh[\frac{\gamma t}{2}(x-\frac{16i\xi}{x})]}{x-16i\xi/x}\Big)-ie^{i\phi}\cosh[\frac{\gamma t}{2}(x+\frac{16i\xi}{x})]\Big(1+\right.\cr\cr
&+&\left.\frac{\tanh[\frac{\gamma t}{2}(x+\frac{16i\xi}{x})]}{x+\frac{16i\xi}{x}}\Big)\right\}\times\sin(\theta).
 \end{eqnarray}
 
In the case of local environments, using Eqs.\eqref{bases} and \eqref{basesopenchain} into \eqref{OBClocal}, we get
\begin{align}
&\dot{a}_{0}(t)=2\gamma[a_{1}(t)+a_{2}(t)+a_{3}(t)],&&\nonumber\\
&\dot{a}_{1}(t)=-2\gamma a_{1}(t)+2\xi a_{10}(t),&&\nonumber\\
&\dot{a}_{2}(t)=-2\gamma a_{2}(t)+2\xi[a_{12}(t)-a_{10}(t)],&&\nonumber\\
&\dot{a}_{3}(t)=-2\gamma a_{3}(t)-2\xi a_{12}(t),&&\nonumber\\
&\dot{a}_{4}(t)=-\gamma a_{4}(t)+\xi a_{7}(t),&&\nonumber\\
&\dot{a}_{5}(t)=-\gamma a_{5}(t)-\xi a_{6}(t),&&\nonumber\\
&\dot{a}_{6}(t)=-\gamma a_{6}(t)+\xi[a_{5}(t)+a_{9}(t)],&&\nonumber\\
&\dot{a}_{7}(t)=-\gamma a_{7}(t)-\xi[a_{4}(t)+a_{8}(t)],&&\nonumber\\
&\dot{a}_{8}(t)=-\gamma a_{8}(t)+\xi a_{7}(t),&&\nonumber\\
&\dot{a}_{9}(t)=-\gamma a_{9}(t)-\xi a_{6}(t),&&\nonumber\\
&\dot{a}_{10}(t)=-2\gamma a_{10}(t)+\xi[a_{2}(t)-a_{1}(t)-a_{11}(t)],&&\nonumber
\end{align}
\begin{align}
&\dot{a}_{11}(t)=-2\gamma a_{11}(t)+\xi[a_{10}(t)-a_{12}(t)],&&\nonumber\\
&\dot{a}_{12}(t)=-2\gamma a_{12}(t)+\xi[a_{11}(t)+a_{3}(t)-a_{2}(t)],&&\nonumber\\
\end{align}
with the initial conditions as Eq.\eqref{initialconditions} and
the following relevant solutions
\begin{align}
&a_{3}(t)=e^{-2\gamma t}\sin^4(t\xi/\sqrt{2})\sin^2(\theta/2),&&\nonumber\\
&a_{8}(t)=-\frac{1}{2}e^{-\gamma t}\sin^{2}(t\xi/\sqrt{2})\sin(\theta)\cos(\phi),&&\nonumber\\
&a_{9}(t)=-\frac{1}{2}e^{-\gamma t}\sin^{2}(t\xi/\sqrt{2})\sin(\theta)\sin(\phi).&&\nonumber\\
\end{align}

\section{Density matrix for closed boundary conditions}\label{appB}
The time evolution of density matrix of a three-qubit chain with closed boundary conditions 
interacting through XY Hamiltonian in presence of quasi-local and local noises are given by Eq.\eqref{CBCchained} and Eq.\eqref{CBClocal}, respectively.
In the case of quasi-local noise, the coefficients of Eq.\eqref{bases} determining the density matrix 
are obtained through the following set of ordinary differential equations:
\begin{align}
&\dot{a}_{0}(t)=2\gamma[a_{1}(t)+a_{2}(t)+a_{3}(t)],&&\nonumber\\
&\dot{a}_{1}(t)=-2\gamma a_{1}(t)+2\xi[a_{11}(t)+a_{13}(t)],&&\nonumber\\
&\dot{a}_{2}(t)=-2\gamma a_{2}(t)+2\xi[a_{15}(t)-a_{11}(t)],&&\nonumber\\
&\dot{a}_{3}(t)=-2\gamma a_{3}(t)-2\xi[a_{13}(t)+a_{15}(t)],&&\nonumber\\
&\dot{a}_{4}(t)=-\gamma a_{4}(t)+\xi[a_{7}(t)+a_{9}(t)],&&\nonumber\\
&\dot{a}_{5}(t)=-\gamma a_{5}(t)-\xi[a_{5}(t)+a_{8}(t)],&&\nonumber\\
&\dot{a}_{6}(t)=-\gamma a_{6}(t)+\xi[a_{5}(t)+a_{9}(t)],&&\nonumber\\
&\dot{a}_{7}(t)=-\gamma a_{7}(t)-\xi[a_{4}(t)+a_{8}(t)],&&\nonumber\\
&\dot{a}_{8}(t)=-\gamma a_{8}(t)+\xi[a_{5}(t)+a_{7}(t)],&&\nonumber\\
&\dot{a}_{9}(t)=-\gamma a_{9}(t)-\xi[a_{4}(t)+a_{6}(t)],&&\nonumber\\
&\dot{a}_{10}(t)=-2\gamma a_{10}(t)+\xi[a_{13}(t)+a_{15}(t)],&&\nonumber\\
&\dot{a}_{11}(t)=-2\gamma a_{11}(t)+\xi[a_{2}(t)+a_{14}(t)-a_{1}(t)-a_{12}(t)],&&\nonumber\\
&\dot{a}_{12}(t)=-2\gamma a_{12}(t)+\xi[a_{11}(t)-a_{15}(t)],&&\nonumber\\
&\dot{a}_{13}(t)=-2\gamma a_{13}(t)+\xi[a_{3}(t)+a_{14}(t)-a_{1}(t)-a_{10}(t)],&&\nonumber\\
&\dot{a}_{14}(t)=-2\gamma a_{14}(t)-\xi[a_{11}(t)+a_{13}(t)],&&\nonumber\\
&\dot{a}_{15}(t)=-2\gamma a_{15}(t)+\xi[a_{3}(t)+a_{12}(t)-a_{2}(t)-a_{10}(t)],&&
\end{align}
where the initial conditions read as Eq.\eqref{initialconditions}.

The coefficients that we need to find the reduced density matrix of the third qubit read
\begin{align}
&a_{3}(t)=\frac{e^{-2\gamma t}}{9}\left\{1+e^{-6\gamma t}-2e^{-3\gamma t}\cos(3t\xi)\right\}\sin^{2}(\theta/2),&&\nonumber\\
 &a_{8}(t)=\frac{e^{-\gamma t}}{6}\left\{e^{-3\gamma t}\cos(\phi-2t\xi)-\cos(\phi+t\xi)\right\}\sin(\theta),&&\nonumber\\
&a_{9}(t)=\frac{e^{-\gamma t}}{6}\left\{e^{-3\gamma t}\sin(\phi-2t\xi)-\sin(\phi+t\xi)\right\}\sin(\theta).&&
 \end{align}
When the chain is affected by local environments, we get
\begin{eqnarray}
\dot{a}_{0}(t)&=&4\gamma[a_{1}(t)+a_{2}(t)+a_{3}(t)+a_{10}(t)+a_{12}(t)+a_{14}(t)],\cr\cr
\dot{a}_{1}(t)&=&-2\gamma[2a_{1}(t)+a_{10}(t)+a_{12}(t)]+2\xi[a_{11}(t)+a_{13}(t)],\cr\cr
\dot{a}_{2}(t)&=&-2\gamma[2a_{1}(t)+a_{10}(t)+a_{14}(t)]-2\xi[a_{11}(t)-a_{15}(t)],\cr\cr
\dot{a}_{3}(t)&=&-2\gamma[2a_{3}(t)+a_{12}(t)+a_{14}(t)]-2\xi[a_{13}(t)+a_{15}(t)],\cr\cr
\dot{a}_{4}(t)&=&-\gamma[2a_{4}(t)+a_{6}(t)+a_{8}(t)]+\xi[a_{7}(t)+a_{9}(t)],\cr\cr
\dot{a}_{5}(t)&=&-\gamma[2a_{5}(t)+a_{7}(t)+a_{9}(t)]-\xi[a_{6}(t)+a_{8}(t)],\cr\cr
\dot{a}_{6}(t)&=&-\gamma[a_{4}(t)+2a_{6}(t)+a_{8}(t)]+\xi[a_{5}(t)+a_{9}(t)],\cr\cr
\dot{a}_{7}(t)&=&-\gamma[a_{5}(t)+2a_{7}(t)+a_{9}(t)]-\xi[a_{4}(t)+a_{8}(t)],\cr\cr
\dot{a}_{8}(t)&=&-\gamma[a_{4}(t)+a_{6}(t)+2a_{8}(t)]+\xi[a_{5}(t)+a_{7}(t)],\cr\cr
\dot{a}_{9}(t)&=&-\gamma[a_{5}(t)+2a_{9}(t)+a_{7}(t)]-\xi[a_{4}(t)+a_{6}(t)],\cr\cr
\dot{a}_{10}(t)&=&-\gamma[a_{1}(t)+a_{2}(t)+4a_{10}(t)+b_{12}(t)+a_{14}(t)]\cr
&+&\xi[a_{13}(t)+a_{15}(t)],\cr\cr
\dot{a}_{11}(t)&=&-\gamma[a_{13}(t)+a_{15}(t)+4a_{11}(t)]\cr
&+&\xi[a_{2}(t)-a_{1}(t)-a_{12}(t)+a_{14}(t)],\cr\cr
\dot{a}_{12}(t)&=&-\gamma[a_{1}(t)+a_{3}(t)+a_{10}(t)+4a_{12}(t)+a_{14}(t)]\cr
&+&\xi[a_{11}(t)-a_{15}(t)],\cr\cr
\dot{a}_{13}(t)&=&-\gamma[a_{11}(t)+4a_{13}(t)+a_{15}(t)]\cr
&+&\xi[a_{14}(t)-a_{1}(t)+a_{3}(t)-a_{10}(t)],\cr\cr
\dot{a}_{14}(t)&=&-\gamma[a_{2}(t)+a_{3}(t)+a_{10}(t)+a_{12}(t)+4a_{14}(t)]\cr
&-&\xi[a_{11}(t)+a_{13}(t)],\cr\cr
\dot{a}_{15}(t)&=&-\gamma[4a_{15}(t)-a_{11}(t)+a_{13}(t)]\cr
&+&\xi[a_{3}(t)-a_{2}(t)-a_{10}(t)+a_{12}(t)],
\end{eqnarray}
with the initial conditions mentioned in Eq.\eqref{initialconditions} and the following relevant solutions:
\begin{align}
&a_{3}(t)=\frac{4}{9}e^{-2\gamma t}\sin^{2}(3t\xi/2)\sin^{2}(\theta/2),&&\nonumber\\
&a_{8}(t)=\frac{1}{3}e^{-\gamma t}\sin(\frac{3t\xi}{2})\sin(\phi-\frac{t\xi}{2})\sin(\theta),&&\nonumber\\
&a_{9}(t)=-\frac{1}{3}e^{-\gamma t}\sin(\frac{3t\xi}{2})\cos(\phi-\frac{t\xi}{2})\sin(\theta).&&
\end{align}

\end{document}